\documentclass [12pt]{article}

\begin{document}
\begin{center}
\textbf{ON A POSSIBLE PHYSICAL METATHEORY OF}
\end{center}

\begin{center}
\textbf{CONSCIOUSNESS}
\end{center}

\begin{center}
Miroljub Dugi\'c$^{1,4}$, Milan M. \'Cirkovi\'c$^{2 }$and Dejan
Rakovi\'c$^{3,4}$
\end{center}

\begin{center}
$^{1}$Department of Physics, Faculty of Science, P.O.Box 60, Kragujevac,
Yugoslavia
\end{center}

\begin{center}
E-mail: \underline {Dugic@knez.uis.kg.ac.yu}
\end{center}

\begin{center}
$^{2}$Astronomical Observatory, Volgina 7, 11160 Belgrade, Yugoslavia
\end{center}

\begin{center}
E-mail: \underline {arioch@eunet.yu}
\end{center}

\begin{center}
$^{3}$Faculty of Electrical Engineering, P.O.Box 35-54, 11000 Belgrade,
Yugoslavia
\end{center}

\begin{center}
E-mail: \underline {Rakovic@net.yu}
\end{center}

\begin{center}
$^{4}$International Anti-Stress Center, Smiljani\'ceva 11/III/7, 11000
Belgrade, Yugoslavia
\end{center}

\begin{center}
E-mail: \underline {info@iasc-bg.org.yu}; WWW address: \underline
{www.iasc-bg.org.yu}
\end{center}

\newpage

\begin{center}
\textbf{ON A POSSIBLE PHYSICAL METATHEORY OF}
\end{center}

\begin{center}
\textbf{CONSCIOUSNESS}
\end{center}

\textbf{Abstract:} We show that the modern quantum mechanics, and
particularly the theory of decoherence, allows formulating a sort
of a physical metatheory of consciousness. Particularly, the
analysis of the necessary conditions for the occurrence of
decoherence, along with the hypothesis that consciousness bears
(more-or-less) well definable physical origin, leads to a wider
physical picture naturally involving consciousness. This can be
considered as a sort of a psycho-physical parallelism, but on
very wide scales bearing some cosmological relevance.

\bigskip

\textbf{1. INTRODUCTION}

\bigskip

In this study we would like to point out that modern quantum
mechanics allows formulating a physical metatheory (metaphysical
theory) of consciousness. This observation comes from some recent
progress in the foundations of the so-called decoherence theory
(Zurek 1991; Dugi\'c 1996, 1997a,b, 1998), as well as some
cosmological discourses (Barrow and Tipler 1986). In addition,
the same program is important in view of the contemporary heated
debate of reductionism versus holism in the philosophy of science
(e.g.\ Edmonds 1999).

We employ practically universally accepted hypothesis in physical
considerations devoted to the issue of consciousness: \textit{there is a physical background (and/or physical basis) of consciousness that, as a physical system, can be described and treated by the methods of the physical sciences}. This partially
trivial assertion will later on prove useful for our considerations, finally
leading to a wider physical picture naturally involving consciousness, and
eventually pointing out something new as regards the connection between
physics and (the physics of) consciousness. As will become clear below, this
reductionist attitude is justified exactly because quantum mechanics (which
we use as a physical basis for discussion) is generally perceived as
introducing a substantial holistic element of modern physics. Therefore, by
pointing out elements necessary for building a metatheory of consciousness,
we may bridge the gap between these two positions, as well as explore the
limits of the theory making process (Landauer 1967).

There is of course no big practical use of the metatheories, generally
speaking. But the observations this way provided usually enrich and/or widen
our point(s) of view. In our opinion, probably the main point of the present
paper is that such a theory---metaphysical theory of
consciousness---naturally follows from the foundations of quantum mechanics.

\bigskip

\textbf{2. A BRIEF ACCOUNT OF THE THEORY OF DECOHERENCE}

\bigskip

Decoherence is a real physical process partly investigated in the
physical laboratories (Devoret et al. 1985a,b; Brune et al. 1996,
Amman et al. 1998). It's history is long and rich of the
different, both conceptual and methodological background.
However, only recently the subject of decoherence met significant
progress and has attracted significant attention of both
theoreticians and experimental physicists.

Decoherence can be qualitatively defined as follows: \textit{it represents a realistic physical process whose effect consists in establishing the (approximate) classical realism for the open physical systems. }Let us briefly discuss
this definition. Quantum mechanics introduces the concept of \textit{quantum indeterminism (quantum uncertainty)}, which
consists in lack of the classical realism for some physical quantities of
the quantum systems. E.g., the following statement does not have sense:
`electron in the hydrogen atom has a definite (relative) position'; rather,
the electron's position is subjected to the famous uncertainty relations of
Heisenberg. As opposite to this, the \textit{classical reality} for a particle's position requires a
definite value of the position in each instant in time generally - being the
particle an isolated system, or in interaction with the surrounding physical
systems. Therefore, existence of definite value of the electron's position
\textit{requires}, in quantum mechanics, an act of measurement of the position.

Therefore, the process of quantum measurement establishes
(D'Espagnat 1971) the classical reality for the measured physical
quantity (quantum-mechanically: observable), and at heart of the
quantum measurement process proves to be (Zurek 1982, Giulini et
al. 1996) the process of decoherence. Needless to say, the
transition from quantum uncertainty to classical reality in the
course of the quantum measurement assumes external intervention
on the measured object (i.e., of the measurement instrument
(apparatus) on the object of measurement), which justifies
referring to the measured object as to an ``open'' quantum
system. In general, by an ``open'' physical system one assumes a
system whose behavior and dynamics (evolution in time) are
substantially determined by its interaction with its environment.

Therefore, one may say that the effect of decoherence establishes at least
approximate classical realism for some of the open system's observables, and
is usually considered (Zurek 1991, Giulini et al. 1996) as the main
candidate for resolving the long standing problem of the ``transition from
quantum to classical'' (Zurek 1991, Omnes 1994).

In the ``macroscopic context'', i.e. as regards the macroscopic physical
systems, the effect of decoherence is expected to meet the following
criteria/requirements: (a) providing a definite border line between the open
quantum system and its environment, (b) establishing at least approximate
classical reality for some observables of the open system as a whole, and
(c) to represent a comparatively quick physical process. Keeping in mind
that the classic-physics world is at the ``macroscopic context'', it is
usually, albeit only plausibly assumed that the process of decoherence
should bear ubiquity and universality in the context of the ``transition
from quantum to classical''.

However, recently (Dugi\'c 1996, 1997a,b), existence of the
necessary conditions for the occurrence of decoherence has been
proved. In particular, it means that the interaction between an
open system (S) and its environment (E) should be of certain kind
as to provide the occurrence of decoherence. As regards the
``microscopic'' physical systems (elementary particles, atoms,
molecules), this result does not mean much; e.g., the interactions
that are not of the kind required are widely used in quantum
mechanics. However, in the ``macroscopic context'', this result
opens some questions.

In the ``macroscopic context'', the occurrence of decoherence is sometimes
(Zurek 1993) plausibly considered as a necessary condition for fulfilling
the above criterion (a), and consequently the other criteria (b) and (c).
Particularly, it can be plausibly stated (Zurek 1993) that the decoherence
provides us with the definite border-line between the two systems, S and E.
Keeping this in mind, one directly concludes that in the ``macroscopic
context'' nonoccurrence of decoherence (as pointed out in Zurek 1993)
\textit{contradicts our macroscopic experience and intuition}. Therefore one may state the question of physical relevance, meaning and
importance of the necessary conditions for the occurrence of decoherence in
the ``macroscopic context''.

\bigskip

\textbf{3. THE ROOTS OF THE PHYSICAL METATHEORY OF CONSCIOUSNESS }

\bigskip

\textit{Prima facie}, the nonoccurrence of decoherence is not physically relevant and can be
interpreted as a pathology of the theory itself\footnote{ More precisely:
one can expect that quantum mechanical formalism, as usually, gives us more
than we can expect and/or interpret in terms of the classical physics, and
particularly in terms of the classical reality. To this end, one may expect
that only those physical models referring to the occurrence of decoherence
could be considered physically relevant or realistic, putting aside the
non-realistic and irrelevant models for which decoherence does not take
place.}. However, this is not really the case. A deeper
physical/interpretational analysis offers an interesting interpretation
naturally involving consciousness.

As it was distinguished in Dugi\'c (1996) and further elaborated
in Dugi\'c, Rakovi\'c and \'Cirkovi\'c (2000), the decoherence
theory allows the following analysis: Let us suppose that the two
systems, an open system S and its environment E are in mutual
interaction not leading to decoherence. Then, according to the
plausible assumption (Zurek 1993) distinguished above, one cannot
determine the border-line between S and E. But suppose that there
exist such coordinate transformations as to allow redefining the
interaction and leading to the definitions of the new physical
systems---the new open system S' and its environment E'. Now,
relative to the coordinates of the new systems, S' and E', one
may say that there occurs the decoherence effect leading to
unambiguous definitions of both systems, S' and E', and
simultaneously defining the desired border-line between the two
systems. This transformation is substantial (cf. Appendix I for
some mathematical details, and for a strict treatment see Dugi\'c
et al 2000), in the sense that the ``old'' systems, S and E,
cannot be even in principle defined or observed.\footnote{ As
regards the whole system (S+E = S'+E'), the canonical
transformations distinguished in Appendix I represent just the
change of representation. However, as to the ``subsystems'', this
change is substantial: it transforms the interaction hamiltonian
from a nonseparable form (for which decoherence does not occur),
to a separable form (for which decoherence can occur). Having in
mind the Zurek's phrase ``no systems-no problem'', we emphasize
that the canonical transformations allow defining a system for
which decoherence may occur---the system S' and its environment
E'---while leaving the ``border line'' between S and E
indefinable.} That is, one deals with the same composite system,
S+E (identical with S'+E'), but the two definitions of the
subsystems (the ``old'' one, S and E, and the ``new'' one, S' and
E') are mutually \textit{exclusive}! The process of decoherence,
which establishes the classical reality only for the ``new''
subsystems, S' and E', clearly states: the open system S' bears
classical reality, and can be defined only simultaneously with its
environment, E'. The composite system cannot be considered
decomposable into the ``old'' ``system'' S and its
``environment'' E: they simply do not bear classical reality,
which is generally expected in the ``macroscopic'' world.

When extended to \textit{complex systems} consisting of a set of
\textit{mutually interacting} (open) macroscopic systems plus
their environments, this notion obtains unexpected element.
Actually, in a set of such systems, the local interaction on one
place determines interaction (and therefore definition of the
systems) at spatially distant place(s), thus making the
macroscopic piece of the Universe (MPU) as an interconnected
physical system, in which definition of each of its part
(element) depends on the definition of a \textit{local} system
and its environment; and this can be rigorously proved (Dugi\'c
1997b; Dugi\'c et al 2000). It cannot be overemphasized: even for
the complex systems, the different definitions of the MPU are
mutually exclusive, in so far as only one of them bears classical
reality.

However, one may ask if the composite system as a whole, can---in
the course of its time evolution---survive transition of the
classical reality from one to another definition\footnote{ Using
the above definitions: the ``new'' system S' and E' looses
classical reality, while the ``old'' systems, now, bear classical
reality.} of the MPU. But this is nonphysical transition, for
\textit{it cannot be observed}. Actually, the conscious observer
could never be aware of this transition, for the simple reason:
according to the assumption (cf. Introduction) that consciousness
bears a macroscopic (Rakovi\'c and Dugi\'c 2000; Dugi\'c and
Rakovi\'c 2000) physical system as its origin, the transformation
from one definition of MPU as a realistic system to another
definition of the MPU bearing classical reality equally refers to
the physical system which is the physical basis of consciousness.
In other words, the different Universes define the different,
mutually exclusive definitions of the systems, which the
consciousness originates from.

This gives us a clue for the physical metatheory of consciousness: The
different definitions of the MPU, bearing classical reality or not, in
principle, define the different consciousness. The physical bases of
consciousness in the different Universes (MPUs) are \textit{mutually exclusive}, bearing the following
substantial characteristics for each Universe: (i) consciousness (through
its macroscopic-physics origin) can be defined only simultaneously with
defining the rest of the MPU, and (ii) the different Universes define the
different, mutually exclusive consciousness.

Therefore, consciousness, treated as a physical system, in the context of
universally valid quantum mechanics is only a relative concept, its physical
characteristics being determined by even remote pieces of the actual
Universe. In practice, it means that observations in a given Universe can be
performed only by the conscious beings physically (in the sense of our
considerations) belonging to that Universe. This, \textit{the relative-metatheory of consciousness }is a sort of
\textit{psycho-physical parallelism} bearing the holistic nature of the physical Universe, which naturally
incorporates consciousness as its part (Wilber 1980; Hoyle 1982; Barrow and
Tipler 1986).

In their lucid and instructive analysis of the collapse problem in quantum
mechanics, Barrow and Tipler (1986, pp. 464-471) offer five basic avenues
for solution. Apart from unattractive solipsism and Everett's
``no-collapse'' theory (which does offer a host of interesting physical and
philosophical issues, but is uninteresting from our present point of view),
these authors suggest that either any being with consciousness can collapse
the wave function by observations, or a ``community'' of such beings can
collectively collapse it, or there is some sort of ``ultimate Observer'' who
is responsible for the collapse. From the point of view exposed above, it is
clear that the nature of MPU and its link with consciousness implies that we
may reject the first option also. Moreover, it could be argued that our
proposal accommodates both the second and the third options.\footnote{
Interestingly enough, these are the two options which---Barrow and Tipler
lament---``have not been explored to any extent'' (Barrow and Tipler 1986,
p. 469). }

In a sense, our suggestion is antithetical to the famous proposal of Eugene
P. Wigner (i.e. Wigner 1967) that the linearity of the Schr\"{o}dinger's
equation fails for conscious entities, and that there is some inherently
non-linear procedure taking place inside those entities. As pointed out by
Penrose (1979, p. 295), this reductionist picture leads to a rather
disturbing view of the reality and actuality of the universe, since
according to this view, by far the largest part of the universe will exist
only as a network of linear superpositions. Our picture, on the contrary,
automatically implies complete realism, even when applied to those parts of
the universe not observed by us, but only implied in the definition of
MPU.\footnote{ This applies to those parts of the universe unobservable by
us \textit{in principle}. For instance, if our universe possesses a particular kind of horizon,
often called event (or de Sitter) horizon, galaxies, stars and possible
intelligent beings beyond this horizon will be unobservable by us \textit{at all times}, both at
present and in arbitrarily distant future. However, our cosmological
theories do suggest that such unobservable galaxies (and an infinite number
of them!) are \textit{real}. } This is certainly a strong merit in the holistic approach
to both quantum mechanics and cosmology.

\bigskip

\textbf{4. SOME COSMOLOGICAL CONSIDERATIONS}

\bigskip

The view that macroscopic parts of our universe play a central
role in physical understanding of consciousness may not be so
surprising \textit{ultimo facie}, especially if one takes
seriously numerous anthropic ``coincidences'' playing a role in
both classical and quantum cosmology (Carter 1974; Barrow and
Tipler 1986). It is a well-known property of the universe that
many of the model parameters in an envisaged complete physical
description must be fine-tuned in order for life and sentience to
be possible; among those are the total cosmological mass density
$\Omega $, magnitude of the cosmological constant $\Lambda $, and
strengths of various couplings (including the celebrated example
of the fine-structure constant $\alpha )$. For instance, it is
well-known that $\Omega $ (\textit{prima facie} a random
variable) has to be in a rather small interval between 0.1 and 10
for life (and, contingently, intelligence and self-awareness) as
we know it to be possible.\footnote{ The total cosmological
density O is a dimensionless quantity defined as the ratio of
actual density of matter (including radiation and any yet-unknown
matter particles and fields) to the critical density necessary
for universe to recollapse under its own gravitational pull.
Therefore, the universe will expand forever if O $ \le $ 1, and
recollapse for O $>$ 1 (which should be taken with the grain of
salt, since the matter fields with ``exotic'' properties may make
the actual situation more complicated). That O is surprisingly
close to unity (within one order of magnitude uncertainty) was
first noted by Dicke in early 1960-ies (famous ``Dicke
coincidences'' from cosmological textbooks). This is the source
of ambiguity which dominated XX century cosmology concerning
future of the universe: will it expand forever, or recollapse to
future singularity of ``Big Crunch''. The best contemporary
evidence strongly suggests the former alternative, although this
can not yet be firmly established. } Other parameters are even
more tightly constrained: it has been argued that a change in
magnitude of nuclear interaction coupling of only about 10{\%}
would make nucleosynthesis of elements necessary for life utterly
impossible.

All these and many other examples testify on the fine-tuning present in the
cosmological initial conditions, i.e. close to the Big Bang singularity. The
similar, although less obvious conclusion applies to the issue of the arrow
of time. As was first discussed by Wiener (1961), the existence of the time
arrow as we perceive it around us is the necessary requisite for
intelligence, and therefore presumably consciousness as well (see also the
discussion in Barrow and Tipler 1986). However, there emerged a sort of
consensus in last several decades on the crucial role played by cosmological
initial conditions in determination of the arrow of time (Penrose 1979;
Price 1996). Initial low-entropy state is a necessary requisite for
subsequent flow of time; however, the real issue then becomes how such
low-entropy state did come into being? The Penrose's estimate of the
probability of spontaneous regularization of Big Bang in order to match the
low-entropy initial conditions---evolving towards the observed state---is
astonishing (see Appendix II for technical details)

\bigskip

\begin{equation}
1 \quad {\rm part \quad in} \quad 10^{10^{123}}(!)
\end{equation}

\bigskip

This stupendous volume of the parameter space \textbf{not} leading to
emergence of intelligence and consciousness cannot fail to emphasize the
highly special nature of the initial cosmological conditions. This result
represents a good basis for our view of the role of cosmological boundary
conditions in the future physical theory of consciousness.

However, this is not the end of the story. The role of MPU in the
considerations above suggests that specific cosmological boundary conditions
of some sort are necessary for the \textit{continuous} existence of consciousness as we know
it. Our conclusion is in accordance with the Empedoclean picture of
contingency between physical and biological processes in the universe (e.g.
Guthrie 1969). One example of such boundary condition is the Wheeler
boundary condition, requiring that intelligent life selects out a single
branch of the universal wave function from ``smeared out'' universe existing
prior to the first measurement interaction. This serves as a physical basis
for Wheeler's so-called Participatory Anthropic Principle, which states that
\textit{observers are necessary to bring universe into being} (cf. Barrow and Tipler 1986). However, even a much weaker assumption could
serve the same purpose in our picture.

As a consequence, one could conjecture that consciousness might
be the essential property of Nature at different structural
levels (macroscopic and microscopic, animate and inanimate), as
widely claimed in traditional esoteric knowledge (Wilber
1980)---which might be supported by analogous mathematical
formalisms of the dynamics of Hopfield's associated neural
networks and Feynman's propagator version of quantum mechanics
(Peru\v{s} 1996)---implying that "collective consciousness" of
Nature itself behaves as a giant nonlocal quantum neural network
with distributed "individual consciousness" processing units.
Such nonlocal pantheistic idea of consciousness is also supported
by Rakovi\'c's physical model of altered and transitional states
of consciousness, explained in the Appendix III. In addition, this
model might provide additional route to the physical solution of
the problem of the wave-packet reduction in the quantum
measurement theory (Rakovi\'c and Dugi\'c 2000).

This picture also offers significant new insights in the nascent field of
physical eschatology---a rather young branch of astrophysics, dealing with
the \textbf{future} fate of astrophysical objects, as well as the universe
itself. Landmark studies in physical eschatology are those of Rees (1969),
Dyson (1979), Tipler (1986) and Adams and Laughlin (1997). Some relevant
issues were also discussed in the monograph of Barrow and Tipler (1986), as
well as several popular-level books (e.g. Davies 1994). Since the
distinction between knowledge in classical cosmology and physical
eschatology depends on the distinction between past and future, several
issues in the physics and philosophy of time are relevant to the assessment
of eschatological results and \textit{vice versa}. In addition, we need to take into account
the almost trivial conclusion, explicitly formulated and defended by Dyson
in his classical paper (Dyson 1979):

It is impossible to calculate in detail the long-range future of the
universe without including the effects of life and intelligence. It is
impossible to calculate the capabilities of life and intelligence without
touching, at least peripherally, philosophical questions. If we are to
examine how intelligent life may be able to guide the physical development
of the universe for its own purposes, we cannot altogether avoid considering
what the values and purposes of intelligent life may be. But as soon as we
mention the words value and purpose, we run into one of the most firmly
entrenched taboos of twentieth-century science.

Future of the universe containing life and intelligence is
\textbf{essentially} different from the past of the same universe in which
there were no such forms of complex organization of matter.\footnote{ In
addition, the premise that MPU generates a particular form of consciousness
immediately obviates the common form of counterfactual cosmological analysis
of (tacitly assumed) ``lifeless'' universes. In this sense our statement
agrees with the abovementioned continual presence of consciousness; while
counterfactuals are essential to theoretical reasoning in physical sciences,
it is crucial that they are \textbf{understood as such}.} Consciousness is
admittedly the most complex such form known, and therefore the issue of the
future of the universe is inseparable from our understanding of the
relationship between consciousness and the (macro)physical world. In
particular, the relative definition of consciousness exposed above is
subject to evolution describing large-scale structure of the universe, as
studied in the eschatological discourse (cf. Tipler 1986).

\bigskip

\textbf{5. DISCUSSION AND PROSPECTS}

\bigskip

The process of decoherence is a realistic (objective) physical
process. It particularly means that no `observer' is required
either for its unfolding, or for the final effect as it is
strongly confirmed by the existing experiments (Devoret et al.
1985a,b; Brune et al. 1996; Amman et al. 1998). Therefore, our
considerations refer to the objective effect of decoherence, and
the above mentioned ``psycho-physical parallelism'' is not the one
introduced by von Neumann and Wigner which assumes substantial
role of consciousness in the process of quantum measurement.
Particularly, in the von Neumann-Wigner interpretation of the
measurement process it is assumed that consciousness is an
external agency necessary and sufficient for the occurrence of
the quantum-mechanical ``collapse'' of the quantum state of the
object of measurement. In our considerations, consciousness is
(cf. Introduction) treated through its physical basis as a
macroscopic, i.e. an open quantum system, thus being a part of
the ``macroscopic piece of the Universe'', not the ``external
agency'' as in the von Neumann-Wigner theory (Wigner 1967).
Therefore, in the context of the universally valid quantum
mechanics (which is precisely the context of the decoherence
theory), one may not expect such a role of consciousness as
regards the ``collapse''. Rather, a new fundamental physical law
is expectable in this regard (Leggett 1980; Prigogine 1997).

Starting point of our considerations is the issue of (non)occurrence of
decoherence (of (non)establishing the classical reality for open quantum
systems). It brought the relative-theory of consciousness. But now, in turn,
one may note that consciousness is able to justify classical reality of the
MPU as well as of the measured quantum observables. Methodologically
speaking, this ``two-direction'' relation between classical reality and
consciousness justifies consistency of our conclusions, and represents the
main characteristic of ``psycho-physical parallelism'' as discussed above.

Interestingly enough, similar holistic thoughts and sentiments have been
expressed (and rather conventionally disregarded) by undoubtedly one of the
greatest physicists of all times, Erwin Schr\"{o}dinger in his 1924. paper
entitled ``Bohr's New Radiation Hypothesis and the Energy Law''. His words,
in the colorful language of those formative years of modern science, sound
appropriate for the conclusion of the present study (Schr\"{o}dinger 1924):

Thus one can also say: a definite stability of the state of the world \textit{sub specie aeternitatis} can
only occur through the \textit{connection} of each individual system with the whole rest of the
world. The separated individual system would be, from the standpoint of the
unity, a chaos. It requires the connection as a permanent \textit{regulator}, without which,
energetically considered, it would wander about at random. Is it an idle
speculation, to find in this a similarity to social, ethical and cultural
phenomena?

\bigskip

{\bf Acknowledgement:} M.M.\'C. wishes to thank Prof. Petar Gruji\'c for his
wholehearted encouragement and support, as well as to Maja Bulatovi\'c and Vesna
Milo\v{s}evi\'c-Zdjelar for kind help in finding several important references.

\bigskip

\textbf{APPENDIX I}

\bigskip

To illustrate the transformations we have in mind (the so-called canonical
transformations) we will use a simplified, unrealistic example (model).

Let us suppose that the (open) system S is defined by its ``coordinate''
$x_S $, while its environment is defined by its ``coordinate'' $x_E $. The
transformations we have in mind are such that define the new (open) system
S' and its environment E' that are defined by their ``coordinates'':

\begin{equation}
\label{eq1}
\begin{array}{l}
 \xi _{S'} = \xi _{S'} (x_S ,x_E ) \\
 \xi _{E'} = \xi _{E'} (x_S ,x_E ) \\
 \end{array}
\end{equation}

\noindent
which states (mathematically) analytical dependence of the ``coordinates''
of the ``new'' systems, S' and E', on the ``coordinates'' of the old
systems, S and E. We further suppose existence of the inverse relations:

\begin{equation}
\label{eq2}
\begin{array}{l}
 x_S = x_S (\xi _{S'} ,'\xi _{E'} ) \\
 x_E = x_E (\xi _{S'} ,\xi _{E'} ) \\
 \end{array}
\end{equation}

\noindent in full analogy with (2). Needless to say, the composite
system is one and the same, i.e. one may state: S+E = S'+E'.

Let us suppose that the interaction in the composite system, when expressed
in terms of ``coordinates'' of the ``old'' systems (S and E) proves not to
lead to decoherence, while when expressed in terms of ``coordinates'' of the
``new'' systems (S' and E'), this interaction leads to decoherence, thus
establishing (approximate) classical reality for the open system S' (and,
simultaneously, for its environment E'). Then the transition:

\begin{equation}
\label{eq3}
(x_S ,x_E )\buildrel {canonicaltransformations} \over \longrightarrow (\xi
_{S'} ,\xi _{E'} )
\end{equation}

\noindent {\it is physically substantial}: due to the lack of
classical reality of the ``old'' system (and its environment), it
states that the inverse transformation to (4) is physically
meaningless. Furthermore, due to the fact that the ``new''
composite system's observables are subject to quantum
uncertainty, the ``old'' composite system' observables are
unobservable.

Actually, the (quantum) measurements of the ``old'' observables, $x_S ,x_E $
require simultaneous measurements of the ``new'' observables $\xi _{S'} ,\xi
_{E'} $ - which is forbidden by the uncertainty relations of the ``new''
observables.

Therefore, even if one may ascribe the (approximate) classical reality to
the ``new'' system's observables, this is not the case with the observables
of the ``old'' systems in so far as the interaction between S and E does not
lead to decoherence.

\bigskip

\textbf{APPENDIX II}

\bigskip

In this Appendix we would like to show how one obtains the most profound
example of fine tuning, the one dealing with the regular nature of the Big
Bang singularity discussed by Penrose (1989) and quoted above in (1).

The relevant measure of gravitational entropy is given, for the case of
black hole, by the famous Bekenstein-Hawking formula (Bekenstein 1973;
Hawking 1975):

\begin{equation}
S_{BH} = \frac{kc^3}{G\hbar }\frac{A}{4}
\end{equation}

\noindent where A is the black hole horizon surface area, and the
rest are fundamental physical constants observed in the universe:
$k$ is the Boltzmann constant, $c$ -- velocity of light, $G$ --
Newtonian gravitational constant, and $\hbar$ is the Planck
constant devided by $2 \pi$. We perceive that the entropy of matter enclosed within
horizon (not entering at all into a physical problem of the state
of such collapsed matter) is proportional to the horizon surface
area. In the simplest case, the one of spherically-symmetric
black hole, horizon is the sphere with area A = 4$\pi $R$^{2}$,
where the relevant radius R is the Schwarschild radius, given as

\begin{equation}
R = \frac{2Gm}{c^2}
\end{equation}

In this formula, $m$ is mass of the black hole (observable, for
instance, through its gravitational attraction of other objects
or even deflection of light rays). Combining this result with (5),
we obtain the following expression for gravitational entropy as a
function of mass:

\begin{equation}
S_{BH} = \frac{2\pi Gk}{\hbar c}m^2
\end{equation}

Now we are in position to calculate the gravitational entropy of the matter
within our visual horizon if we could somehow collapse it into a single
gigantic black hole (since we can reasonably estimate the mass of all matter
within the horizon). It will be the \textit{maximal} entropy state, since, as shown by
Bekenstein, the state of matter in the black hole is the most probable one,
as far as gravitational interaction is concerned. Any other state (for
instance the one we observe at present, where matter is clumped in galaxies,
stars, planets, etc. and there is just a small number of black holes) is a
priori \textit{less probable}. It is worth noticing, however, that in the context of contemporary
relativistic cosmological models, such situation actually occurred in the
past: all matter within our visual horizon today was within the initial Big
Bang singularity, which has much in common with the (local) black hole
singularities.

Finally, to establish quantitatively how much less probable is
the observed entropy in comparison to (7), we need the
historically all-important Shannon formula, giving the
relationship between entropy and information (Shannon 1948):

\begin{equation}
S = - k\sum\limits_i {p_i \ln p_i ,}
\end{equation}

\noindent where $p_{i}$ is the probability of system considered
being in state $i$.\footnote{ It is worth noticing that there is
some ambiguity in the literature, which concept of entropy is the
more ``fundamental'' one and therefore (8) is sometimes written
without the Boltzmann constant $k$, and it is said that the
Shannon entropy for thermodynamical systems is equal to the
fine-grained entropy in units of $k$. We neglect this rather
semantic issue in the present discussion. } Even on a qualitative
level, it is clear that the presence of logarithm in (8) is the
source of huge exponential terms such as the one in eq. (1). It
gives us a proper lever to compare our universe with the case in
which all matter within our horizon is located in black holes. It
comes out that, as Penrose (1989) discusses, our universe is of
so small entropy compared to the generic one, that the
probability of its reaching the observed state is stupendously
small, as in (1). This is, as correctly emphasized by Penrose and
Price, not only a source of cosmological, but probably all other
arrows of time as well, and a profound example of fine-tuning to
be accounted for by unified field theories of the near future.

\bigskip

\textbf{APPENDIX III}

\bigskip

The goal of this Appendix is to qualitatively sketch the model of
transitional states of consciousness developed by Rakovi\'c (1995;
Rakovi\'c et al. 2000). Namely, these states might be deeply
connected with the role of "collective consciousness" (as a
composite quantum state $\Phi $ of all "individual consciousness"
$\phi _k $: $\Phi \sim \prod\limits_k {\phi _k } )$ in the
quantum theory of measurement, where "collective consciousness"
with its assembling (equivalent to convergence of Feynman's
propagator quantum mechanics to one of its propagators, $\Phi _i
)$ contributes in channeling reduction of initial wave function
$\Psi $ into one of (possible) probabilistic eigenstates $\Psi _i
$ - which implies that "collapse" could be related with
generation of microparticles' local wormholes in highly
noninertial microparticle's interactions in quantum measurement
situations (fully equivalent to extremely strong gravitational
fields according to Einstein's Principle of equivalence, where
relativistic generation of wormholes is predicted; cf. Morris,
Thorne and Yurtsever 1988; Thorne 1994). In a similar vein, in
the Penrose's gravitationally induced collapse (e.g. Penrose
1994) the very mechanism for this process could be continuous
opening and closing of local microparticle's wormholes, addresses
of their exits being related (probabilistically) to one of
(possible) eigenstates $\Psi _i $ of corresponding quantum
system---and everything being related to corresponding
(probabilistic) assembling $\Phi $\textbf{$ \to $}$\Phi _i $ of
"collective consciousness", thus channeling the collapse $\Psi
$\textbf{$ \to $}$\Psi _i $\textbf{.}

The question how it is possible that these highly noninertial
microparticles' processes with inevitable relativistic generation of
microparticles' wormholes and other envisaged quantum-gravitational effects
were not taken into account within quantum mechanics which is yet extremely
accurate theory---might be answered as it was, but within the ad hoc von
Neumann's "projection postulate" (von Neumann 1955) to account for
quantum-mechanical "wave packet collapse" in quantum measurement situations
(implying also that this \textit{ad hoc} procedure is based on quantum gravitational
phenomena, as suggested by Penrose, being on deeper physical level than
quantum mechanical ones!). On the other hand, the nonlocality of usually
conceived "collective consciousness" provides additional evidence for the
nonlocal nature of quantum mechanics---otherwise demonstrated by tests of
Bell's inequalities and the Einstein-Podolsky-Rosen effect (Bell 1987;
Aspect, Dalibard, and Roger 1982).

It should be also pointed out that the above "collective
consciousness"' assembling $\Phi _i  \quad (i = 1,2,3...)$ in
quantum theory of measurement should be interpreted as purely
probabilistic (with relative frequency of their appearance given
by quantum-mechanical probability $\left| {a_i } \right|^2$ of
realization of corresponding microparticles' eigenstates $\Psi _i
$, where $\Phi \Psi = \sum\limits_i {a_i \Phi _i \Psi _i } )$,
depending not on the previous history of the repeatedly prepared
quantum system. However, this might not be the case for
biological "individual consciousness"' assembling, being
history-dependent deterministic one (resulting in deterministic
convergence of the consciousness-related-acupuncture
electromagnetic/ionic microwave ultra-low frequency-modulated
oscillatory holographic Hopfield-like associative neural network
to the particular attractor in the potential hypersurface
(Jovanovi\'c-Ignjati\'c and Rakovi\'c 1999; Rakovi\'c et al.
2000), or equivalently to deterministic convergence of Feynman's
propagator quantum mechanics to the particular propagator
corresponding to $\phi _k $, fixedly determined by "individual
consciousness"), implying that strong preferences in individual
futures might exist, governed by individual mental loads, as
widely claimed in Eastern tradition (Wilber 1980; Vuji\v{c}in
1996). The same may apply to collective futures $\Phi _i $, also
governed by interpersonal mental loads (Rakovi\'c 2000). It
should be also noted that these preferences in individual and
collective futures might be anticipated in transitional states of
consciousness that might be the basis of intuition, precognition
and deep creative insights (Jahn 1982). What is really
anticipated in transitional states of consciousness of
"individual consciousness" might be the evolved state of cosmic
"collective consciousness" $\Phi $(t) (to which our "individual
consciousness" $\phi _k $ has access, being the presumed
constituting part of cosmic "collective consciousness"), which is
quantum-mechanically described by deterministic unitary evolution
governed by the Schr\"{o}dinger equation.

A hypothesis that nonlocal individual/collective consciousness re-assembling
($\Phi $\textbf{$ \to $}$\Phi _i )$ is possible, with direct influence on
the collapse of the observed system ($\Psi $\textbf{$ \to $}$\Psi _i )$,
might be also supported by Princeton PEAR human/machine experiments (Jahn
and Dunne 1988), where (even distant) human operators, solely by volition,
have been able to influence the sophisticated machines with (otherwise)
strictly random outputs, in a statistically repeatable effects (of the order
of a few parts in ten thousand) - but individually not reproducible at any
moment, which is a standard request in contemporary scientific experiments.
All this can be accounted by intentional transitional transpersonal
biological (non-Schr\"{o}dinger governed) quantum gravitational tunneling of
the "operator's individual consciousness" with mental addressing on the
"machine's content of collective consciousness", channeling intentionally
the "operator/machine composite state of collective consciousness" ($\Phi
$\textbf{$ \to $}$\Phi _i )$, thus automatically influencing the machine
output ($\Psi $\textbf{$ \to $}$\Psi _i )$ in the non-Schr\"{o}dinger
quantum-gravitationally governed collapse-like process ($\Phi \Psi
$\textbf{$ \to $}$\Phi _i \Psi _i )$. As a consequence one could further
support the conjecture that consciousness might be essential property of
Nature at different structural levels, macroscopic and microscopic, animate
and inanimate, being presumably related to the unified field itself (Hagelin
1987).

\bigskip

\textbf{REFERENCES}

\bigskip

Adams, F. C. and Laughlin, G. 1997, \textit{Rev. Mod. Phys.} \textbf{69}, 337.

Amman H. et al. 1998, \textit{Phys. Rev. Lett.} \textbf{80}, 4111.

Aspect, A., Dalibard, J. and Roger, G. 1982, \textit{Phys. Rev. Lett. }\textbf{49}, 1804.

Barrow, J. D. and Tipler, F. J. 1986, The Anthropic Cosmological Principle
(Oxford University Press, New York).

Bekenstein, J. D. 1973, \textit{Phys. Rev. D} \textbf{7}, 2333.

Bell, J. S. 1987, Speakable and Unspeakable in Quantum Mechanics (Cambridge
University Press, Cambridge).

Brune M. et al. 1996, \textit{Phys. Rev. Lett.} \textbf{77}, 4887.

Carter, B. 1974, in Physical Cosmology and Philosophy, ed. by Leslie, J.
(1990, Macmillan, London), 131.

\'Cirkovi\'c, M. M. 2001, \textit{Studies in History and
Philosophy of Modern Physics}, submitted for publication.

\'Cirkovi\'c, M. M. and Bostrom, N. 2000, \textit{Astrophys. Space
Sci.} \textbf{274}, 675.

Davies, P. C. W. 1994, The Last Three Minutes (Basic Books, New York).

D'Espagnat B. 1971, Conceptual Foundations of Quantum Mechanics (Benjamin,
Reading, MA).

Devoret M.H. et al. 1985a, \textit{Phys. Rev. Lett.} \textbf{55}, 1543.

Devoret M.H. et al. 1985b, \textit{Phys. Rev. Lett.} \textbf{55}, 1908.

Dugi\'c, M. 1996, \textit{Physica Scripta} \textbf{53}, 9.

Dugi\'c, M. 1997a, \textit{Physica Scripta} \textbf{56}, 560.

Dugi\'c, M. 1997b, A Contribution to the Foundations of the Theory
of Decoherence in Nonrelativistic Quantum Mechanics, Ph.D. Thesis,
Faculty of Science, University of Kragujevac (in Serbian).

Dugi\'c M. 1998, \textit{J. Res. Phys.} \textbf{27}, 141.

Dugi\'c, M. and Rakovi\'c, D. 2000, \textit{Eur. Phys. J. B}
\textbf{13}, 781.

Dugi\'c, M., Rakovi\'c, D., and \'Cirkovi\'c, M. M. 2000, I.
Kononenko, ed., Proceedings of New Science of Consciousness
(Information Society, Ljubljana).

Dyson, F. 1979, \textit{Rev. Mod. Phys. }\textbf{51}, 447.

Edmonds, R. 1999, \textit{Foundations of Science} \textbf{4}, 57.

Giulini, D., Joos, E., Kiefer, C., Kupsch, J., Stamatescu, I.-O. and Zeh, H.
D. 1996, Decoherence and the Appearance of a Classical World in Quantum
Theory (Springer, Berlin).

Guthrie, W. K. C. 1969, \textit{A History of Greek Philosophy II} (Cambridge University Press, London).

Hagelin, J. S. 1987, \textit{Modern Sci. {\&} Vedic Sci.} \textbf{1}, 29.

Hawking, S. W. 1975, \textit{Commun. Math. Phys.} \textbf{43}, 199.

Hoyle, F. 1982, \textit{Ann. Rev. Astron. Astrophys. }\textbf{20}, 1.

Jahn, R. G. 1982, \textit{Proc. IEEE }\textbf{70}, 136.

Jahn, J. R. and Dunne, B. J. 1988, Margins of Reality (Harcourt
Brace, New York).\footnote{ In connection with this reference, see
also many archival publications and technical reports by PEAR (Princeton
Engineering Anomalies Research); relevant pointers are on the following WWW
address: www.princeton.edu/$\sim $rdnelson/pear.html.}

Jovanovi\'c-Ignjati\'c, Z. and Rakovi\'c, D. 1999, \textit{Acup.
{\&} Electro-Therap. Res. Int. J. }\textbf{24}, 105

Landauer, R. 1967, \textit{IEEE Spectrum}, September, 105.

Leggett A.J. 1980, \textit{Prog. Theor Phys. Suppl.} \textbf{64}, 80.

Morris, M. S., Thorne, K. S. and Yurtsever, U. 1988, \textit{Phys. Rev. Lett. }\textbf{61}, 1446.

Omnes, R. 1994, The Interpretation of Quantum Mechanics (Princeton
University Press, Princeton).

Penrose, R. 1979, in General Relativity: An Einstein Centenary, ed. by
Hawking, S. W. and Israel, W. (Cambridge University Press, Cambridge), 581.

Penrose, R. 1989, The Emperor's New Mind (Oxford University Press, Oxford).

Penrose, R. 1994, Shadows of the Mind. A Search for the Missing Science of
Consciousness (Oxford University Press, Oxford, 1994).

Peru\v{s}, M. 1996, \textit{Informatica} \textbf{20}, 173.

Prigogine I. 1997, The End of Certainty (The Free Press, New York).

Price, H. 1996, Time's Arrow and Archimedes' Point (Oxford University Press,
Oxford).

Rakovi\'c, D. 1995, in D. Rakovi\'c and Dj. Koruga, eds.,
Consciousness: Scientific Challenge of the 21st Century (ECPD,
Belgrade).

Rakovi\'c, D. 2000, \textit{Int. J. Appl. Sci. {\&} Computations}
\textbf{7}, 174.

Rakovi\'c, D. et al. 2000, \textit{Electro- and Magnetobiology
}\textbf{19}, 195.

Rakovi\'c, D. and Dugi\'c, M. 2000, \textit{Informatica}, in
press.

Rees, M. J. 1969, \textit{Observatory }\textbf{89}, 193.

Schr\"{o}dinger, E. 1924, \textit{Naturwiss. }\textbf{12}, 720.

Shannon, C. E. 1948, \textit{The Bell System Technical Journal }\textbf{27}, 379.

Thorne, K. S. 1994, Black Holes and Time Warps: Einstein's Outrageous Legacy
(Picador, London).

Tipler, F. J. 1986, \textit{Int. J. Theor. Phys.} \textbf{25}, 617.

\noindent
von Neumann, J. 1955, Mathematical Foundations of Quantum Mechanics
(Princeton University Press, Princeton).

Vuji\v{c}in, P. 1995, in D. Rakovi\'c and Dj. Koruga, eds.,
Consciousness: Scientific Challenge of 21st Century (ECPD,
Belgrade).

Wigner, E. P. 1967, Symmetries and Reflections (Indiana University Press,
Bloomington).

Wilber, K. 1980, The Atman Project (Quest, Wheaton, IL).

Zurek, W. H. 1982, \textit{Phys. Rev. D} \textbf{26}, 1862.

Zurek, W. H. 1991, \textit{Physics Today} \textbf{48}, 36.

Zurek, W. H. 1993, \textit{Prog. Theor. Phys.} \textbf{89}, 281.

\end{document}